\documentclass[preprintnumbers,amsmath,amssymbm,prd]{revtex4}
\usepackage{epsfig}
\usepackage{graphicx}
\usepackage{amssymb}

\begin{document}
\title{No hair for spherically symmetric neutral black holes: nonminimally coupled massive scalar fields}
\author{Shahar Hod}
\affiliation{The Ruppin Academic Center, Emeq Hefer 40250, Israel}
\affiliation{ }
\affiliation{The Hadassah Institute, Jerusalem 91010, Israel}
\date{\today}

\begin{abstract}
\ \ \ It is proved that spherically symmetric asymptotically flat
neutral black holes cannot support spatially regular static
configurations made of massive scalar fields with non-minimal
coupling to gravity. Interestingly, our compact no-hair theorem is
valid for generic values of the dimensionless physical parameter
$\xi$ which quantifies the strength of coupling between the scalar
field and the spacetime curvature.
\end{abstract}
\bigskip
\maketitle

\section{Introduction}

The interplay between classical black-hole spacetimes and scalar
matter fields has attracted much attention from physicists and
mathematicians during the last five decades. Interestingly, early
investigations of the non-linearly coupled Einstein-scalar field
equations made by Bekenstein \cite{Bek1} and Teitelboim \cite{Teit}
(see also \cite{Bekto,Chas,Heu,Bek2,BekMay,Bek20} and references
therein) have revealed that spatially regular static matter
configurations made of massive scalar fields with minimal coupling
to gravity cannot be supported in asymptotically flat regular
\cite{Noteregn} black-hole spacetimes \cite{Noteev}. [It is worth
noting that, as opposed to the {\it static} scalar fields considered
in the physically important early no-hair theorems of \cite{Bek1}
and \cite{Teit}, {\it stationary} massive scalar (and, in general,
bosonic) fields can be supported in asymptotically flat rotating
black-hole spacetimes \cite{Hodrc,Herkr,NR1}].

Interestingly, the rigorous derivation of an analogous no-hair
theorem for the more generic case of black holes interacting with
spatially regular scalar fields which are characterized by {\it
non}-minimal coupling to the spacetime curvature turns out to be a
mathematically challenging task. In particular, the physically
influential and mathematically elegant theorems presented by
Bekenstein and Mayo \cite{BekMay,Bek20} imply that spherically
symmetric asymptotically flat black holes cannot support
nonminimally coupled spatially regular neutral scalar fields in the
restricted regimes $\xi<0$ and $\xi\geq1/2$ of the dimensionless
field-curvature non-minimal coupling parameter $\xi$
\cite{Noteev,Notewx,Noters,Hodrs1,Hodrs2,NRs,NR2,NR3}. (It is worth
emphasizing that Bekenstein and Mayo \cite{BekMay} have also
presented an extensive study of the no-hair property for
non-minimally coupled, complex, potentially charged scalar fields in
the composed Einstein-Maxwell theory).

The no-hair theorem presented in \cite{Hodrs1}, which is valid for
generic inner boundary conditions, can be used to exclude the
existence of spherically symmetric asymptotically flat static hairy
black-hole configurations supporting massive scalar fields with
nonminimal coupling to gravity in the dimensionless regimes $\xi<0$
and $\xi>1/4$.

To the best of our knowledge, no mathematically rigorous no-hair
theorem has thus far been presented in the physical literature which
rules out the possible existence of static hairy black-hole
configurations with regular event horizons which are made of
nonminimally coupled massive neutral scalar fields with a
field-curvature coupling parameter in the dimensionless physical
regime $0<\xi\leq1/4$ \cite{BekMay,Bek20}.

The main goal of the present paper is to present a compact no-hair
theorem which explicitly proves that spherically symmetric
asymptotically flat neutral black holes with regular event horizons
cannot support static matter configurations made of massive scalar
fields with non-minimal coupling to gravity. Interestingly, our
novel no-hair theorem, to be presented below, is valid for {\it
generic} values of the physical parameter $\xi$ which characterizes
the non-minimal field-curvature coupling in the composed
black-hole-massive-scalar-field system (In particular, in the
present paper we shall extend the important no-hair theorems of
\cite{BekMay,Bek20,Hodrs1} to the regime of nonminimally coupled
massive neutral scalar fields with $0<\xi\leq1/4$).

\section{Description of the system}

We consider a spherically symmetric static matter distribution
described by a massive scalar field $\psi$ with nonminimal coupling
to gravity which is non-linearly coupled to a neutral black hole of
horizon radius $r_{\text{H}}$. The spherically symmetric static
curved spacetime of the composed black-hole-massive-scalar-field
system is characterized by the radially-dependent line element
\cite{BekMay} (we shall use natural units in which $G=c=\hbar=1$)
\begin{equation}\label{Eq1}
ds^2=-e^{\nu}dt^2+e^{\lambda}dr^2+r^2(d\theta^2+\sin^2\theta
d\phi^2)\ ,
\end{equation}
where $\{t,r,\theta,\phi\}$ are the Schwarzschild spacetime
coordinates and $\{\nu=\nu(r),\lambda=\lambda(r)\}$. As proved in
\cite{BekMay}, non-extremal black-hole spacetimes with regular event
horizons \cite{Noteregn} are characterized by the simple
near-horizon functional relations
\begin{equation}\label{Eq2}
e^{-\lambda}=L\cdot x+O(x^2)\ \ \ \ \text{where}\ \ \ \ x\equiv
{{r-r_{\text{H}}}\over{r_{\text{H}}}}\ \ \ ; \ \ \ L>0\
\end{equation}
[as explicitly shown in \cite{BekMay}, the expansion coefficient in
(\ref{Eq2}) is given by $L\equiv 1+8\pi T^{t}_{t}(r_{\text{H}})
r^2_{\text{H}}>0$, where $-T^{t}_{t}(r_{\text{H}})$ is the energy
density which characterizes the matter fields at the horizon
$r=r_{\text{H}}$ of the black-hole spacetime], and
\begin{equation}\label{Eq3}
\lambda{'}r_{\text{H}}=-{{1}\over{x}}+O(1)\ \ \ \ ; \ \ \ \
\nu{'}r_{\text{H}}={{1}\over{x}}+O(1)\  ,
\end{equation}
where a prime ${'}$ denotes a spatial derivative with respect to the
radial coordinate $r$. In addition, asymptotically flat spacetimes
are characterized by the large-$r$ functional relations
\cite{BekMay}
\begin{equation}\label{Eq4}
\nu\sim M/r\ \ \ \text{and}\ \ \ \lambda\sim M/r\ \ \ \ \text{for}\
\ \ \ r\to\infty\  ,
\end{equation}
where $M$ is the total ADM mass (as measured by asymptotic
observers) of the hairy black-hole spacetime.

The non-minimally coupled scalar field $\psi$ of mass $\mu$ [note
that the mass parameter $\mu$ of the scalar field stands for
$\mu/\hbar$ and therefore has the dimensions of (length)$^{-1}$] is
characterized by the action \cite{BekMay}
\begin{equation}\label{Eq5}
S=S_{EH}-{1\over2}\int\big(\partial_{\alpha}\psi\partial^{\alpha}\psi+\mu^2\psi^2+\xi
R\psi^2\big)\sqrt{-g}d^4x\ ,
\end{equation}
where $S_{\text{EH}}$ denotes the Einstein-Hilbert action of the
curved spacetime and the dimensionless physical parameter $\xi$
quantifies the strength of the nonminimal coupling between the
scalar field $\psi$ and the scalar curvature $R(r)$ [see Eq.
(\ref{Eq23}) below] of the black-hole spacetime. It is worth noting
that the no-hair theorem for scalar-tensor theories presented in
\cite{NRs}, which is based on a simple transformation of the action
from the Jordan frame to the Einstein frame, is not applicable to
the action (\ref{Eq5}) considered in the present paper. In
particular, the scalar-curvature coupling considered in \cite{NRs}
is proportional to $\psi R$, whereas the action (\ref{Eq5}) for the
massive scalar field is characterized by a nonminimal
scalar-curvature coupling of the functional form $\psi^2 R$.

Note that an asymptotically flat spacetime is characterized by the
simple functional relation \cite{BekMay}
\begin{equation}\label{Eq6}
R(r\to\infty)\to 0\  .
\end{equation}
In addition, as discussed in \cite{BekMay}, physically acceptable
spacetimes are characterized by finite and positive values of the
effective asymptotic ($r/M\gg1$) gravitational constant
$G_{\text{eff}}=G[1-8\pi G\xi\psi^2(r/M\gg1)]$ \cite{BekMay}. This
physical requirement implies that the radial eigenfunction of the
nonminimally coupled massive scalar fields is bounded asymptotically
by the simple relations
\begin{equation}\label{Eq7}
-\infty<8\pi\xi\psi^2<1\ \ \ \ \text{for}\ \ \ \ r/M\gg1\  .
\end{equation}

The action (\ref{Eq5}) of the nonminimally coupled massive scalar
field yields the differential equation \cite{BekMay}
\begin{equation}\label{Eq8}
\partial_{\alpha}\partial^{\alpha}\psi-(\mu^2+\xi R)\psi=0\  ,
\end{equation}
which, in the spherically symmetric static black-hole spacetime
(\ref{Eq1}) and for a real scalar field $\psi=\psi(r)$ whose spatial
behavior depends only on the radial coordinate $r$, can be written
in the form
\begin{equation}\label{Eq9}
\psi{''}+{1\over2}\big({{4}\over{r}}+\nu{'}-\lambda{'}\big)\psi{'}-(\mu^2+\xi
R) e^{\lambda}\psi=0\  .
\end{equation}
[It is worth emphasizing that the radial differential equation
(\ref{Eq9}) is valid for a real and static scalar field which
inherits the symmetries of the spherically symmetric static
black-hole spacetime (\ref{Eq1})].

The characteristic action (\ref{Eq5}) of the nonminimally coupled
massive scalar field also yields the following functional
expressions \cite{BekMay}
\begin{equation}\label{Eq10}
T^{t}_{t}=e^{-\lambda}{{\xi(4/r-\lambda{'})\psi\psi{'}+(2\xi-1/2)(\psi{'})^2+2\xi\psi\psi{''}}
\over{1-8\pi\xi\psi^2}}-{{\mu^2\psi^2}\over{2(1-8\pi\xi\psi^2)}}\ ,
\end{equation}
\begin{equation}\label{Eq11}
T^{t}_{t}-T^{r}_{r}=e^{-\lambda}{{(2\xi-1)(\psi{'})^2+2\xi\psi\psi{''}-\xi(\nu+\lambda){'}\psi\psi{'}}
\over{1-8\pi\xi\psi^2}}\  ,
\end{equation}
and
\begin{equation}\label{Eq12}
T^{t}_{t}-T^{\phi}_{\phi}=e^{-\lambda}{{\xi(2/r-\nu{'})\psi\psi{'}}
\over{1-8\pi\xi\psi^2}}
\end{equation}
for the components of the energy-momentum tensor [It is worth noting
that, except of the Einstein relation $R=-8\pi T$, our no-hair
theorem, to be presented below, does not rely on the explicit forms
of the Einstein field equations $G^{\mu}_{\nu}=8\pi T^{\mu}_{\nu}$.
Nevertheless, for the interested readers, we note that the
corresponding functional expressions of the Einstein tensor
components $G^{\mu}_{\nu}$ can be found, for example, in Eqs. (2.4),
(2.5) and (2.6) of \cite{Asp}]. As discussed in \cite{BekMay},
causality requirements imply that, for physically acceptable
systems, the components of the energy-momentum tensor are
characterized by the simple inequalities
\begin{equation}\label{Eq13}
|T^{\theta}_{\theta}|=|T^{\phi}_{\phi}|\leq|T^{t}_{t}|\geq|T^{r}_{r}|\
.
\end{equation}
In particular, below we shall use the following functional relations
\cite{BekMay}
\begin{equation}\label{Eq14}
\text{sgn}(T^{t}_{t})=\text{sgn}(T^{t}_{t}-T^{r}_{r})=\text{sgn}(T^{t}_{t}-T^{\phi}_{\phi})\
,
\end{equation}
which provide necessary conditions for the validity of the energy
conditions (\ref{Eq13}) derived in \cite{BekMay} for physically
acceptable systems. Finally, following \cite{BekMay} we stress the
important fact that physically acceptable spacetimes are also
characterized by energy-momentum tensors with finite mixed
components \cite{BekMay}:
\begin{equation}\label{Eq15}
\{|T^{t}_{t}|,|T^{r}_{r}|,|T^{\theta}_{\theta}|,|T^{\phi}_{\phi}|\}<\infty\
.
\end{equation}

\section{Review of former analytical results: No nonminimally coupled
massive scalar hair in the regimes $\xi<0$ and $\xi>{1\over 4}$}

Before we proceed, it is important to mention that the no-hair
theorem presented in \cite{Hodrs1} rules out the existence of
asymptotically flat static hairy black-hole configurations with
regular event horizons supporting massive scalar fields with
nonminimal coupling to gravity in the dimensionless regimes $\xi<0$
and $\xi>1/4$. (It is worth stressing the fact that the
no-scalar-hair theorem presented in \cite{Hodrs1} is valid for {\it
generic} inner boundary conditions. In particular, this theorem is
valid for both asymptotically flat black-hole spacetimes with
absorbing horizons and for asymptotically flat stars with compact
reflecting boundaries \cite{Hodrs1}). For completeness of the
presentation, in the present section we shall give a brief sketch of
the no-hair theorem presented in \cite{Hodrs1}.

Taking cognizance of the characteristic asymptotic functional
relations (\ref{Eq4}) and (\ref{Eq6}) of the static black-hole
spacetime (\ref{Eq1}), one finds from (\ref{Eq9}) the mathematically
compact radial scalar equation \cite{Hodrs1}
\begin{equation}\label{Eq16}
\psi{''}+{{2}\over{r}}\psi{'}-\mu^2\psi=0\ \ \ \ \text{for}\ \ \ \
r/M\gg1\  ,
\end{equation}
which determines the spatial behavior of the massive scalar fields
in the asymptotic $r\gg M$ region. The physically acceptable
solution of the radial differential equation (\ref{Eq16}) which
respects the characteristic asymptotic relation (\ref{Eq7})
\cite{BekMay} is given by
\begin{equation}\label{Eq17}
\psi(r)=A\cdot{{e^{-\mu r}}\over{\mu r}}\ \ \ \ \text{for}\ \ \ \
r/M\gg1\ ,
\end{equation}
where $A$ is a dimensionless normalization constant.

Taking cognizance of Eqs. (\ref{Eq4}), (\ref{Eq10}), (\ref{Eq12}),
and (\ref{Eq17}), one obtains the asymptotic functional expressions
\cite{Hodrs1}
\begin{equation}\label{Eq18}
T^{t}_{t}=(4\xi-1)\mu^2\psi^2\cdot[1+O(M/r)]\
\end{equation}
and
\begin{equation}\label{Eq19}
T^{t}_{t}-T^{\phi}_{\phi}=-\xi{{2\mu}\over{r}}\psi^2\cdot[1+O(M/r)]\
\end{equation}
for the energy-momentum components of the spatially regular
nonminimally coupled massive scalar fields. In particular, the
expressions (\ref{Eq18}) and (\ref{Eq19}) yield the mathematically
simple and physically important asymptotic relation
\begin{equation}\label{Eq20}
\text{sgn}(T^{t}_{t})=-\text{sgn}(T^{t}_{t}-T^{\phi}_{\phi})\ \ \ \
\text{for}\ \ \ \ \xi<0 \ \ \text{or}\ \ \xi>1/4\  .
\end{equation}

Interestingly, the relation (\ref{Eq20}) between the components of
the energy-momentum tensor which characterizes the nonminimally
coupled massive scalar field, violates the fundamental relation
(\ref{Eq14}) which, as explicitly proved in \cite{BekMay},
characterizes physically acceptable Einstein-matter systems. One
therefore realizes that spatially regular static matter
configurations made of massive scalar fields nonminimally coupled to
gravity in the regimes $\xi\leq0$ [note that the early no-hair
theorems of \cite{Bek1} and \cite{Teit} explicitly rule out the
existence of external matter configurations (`hair') which are made
of spatially regular minimally coupled ($\xi=0$) static massive
scalar fields] and $\xi>1/4$ cannot be supported in spherically
symmetric asymptotically flat neutral black-hole spacetimes
\cite{Hodrs1}.

\section{The no-hair theorem for the nonminimally coupled massive scalar fields in the regime $\xi>0$}

In the present section we shall complete our no-hair theorem for the
composed black-hole-scalar-field system by explicitly proving that
static matter configurations made of spatially regular nonminimally
coupled massive scalar fields in the dimensionless physical regime
$\xi>0$ {\it cannot} be supported in spherically symmetric
asymptotically flat neutral black-hole spacetimes.

We shall first prove that the characteristic eigenfunction $\psi(r)$
of the nonminimally coupled self-gravitating massive scalar fields
is a non-monotonic function of the radial coordinate $r$. Taking
cognizance of the characteristic asymptotic behavior [see Eq.
(\ref{Eq17})]
\begin{equation}\label{Eq21}
\psi(r/M\to\infty)\to0\
\end{equation}
of the spatially regular massive scalar fields, one deduces that if
the radial scalar eigenfunction is characterized by the near-horizon
behavior $\psi(x\to0)\to0$ then it must possess, in accord with the
above mentioned claim, an extremum point in the exterior black-hole
spacetime. We shall now prove that scalar eigenfunctions with the
near-horizon behavior
\begin{equation}\label{Eq22}
\psi(x\to0)\neq0\
\end{equation}
are also characterized by a non-monotonic radial profile.

Taking cognizance of Eqs. (\ref{Eq10}), (\ref{Eq11}), and
(\ref{Eq12}), and using the Einstein relation $R=-8\pi T$ [here the
radially-dependent function $T(r)$ denotes the trace of the
energy-momentum tensor which characterizes the matter fields], one
finds the functional expression
\begin{equation}\label{Eq23}
R=-{{8\pi}\over{1-8\pi\xi\psi^2}}\Big\{e^{-\lambda}\Big[\xi\big({{12}\over{r}}+3\nu{'}-3\lambda{'}\big)\psi\psi{'}+
6\xi\psi\psi{''}+(6\xi-1)(\psi{'})^2\Big]-2\mu^2\psi^2\Big\}\
\end{equation}
for the Ricci scalar curvature which characterizes the spherically
symmetric composed black-hole-massive-scalar-field system.
Substituting (\ref{Eq23}) into (\ref{Eq9}), one obtains the (rather
cumbersome) radial differential equation
\begin{equation}\label{Eq24}
\psi{''}\cdot\big[1+8\pi\xi(6\xi-1)\psi^2\big]
+\psi{'}\cdot\Big[{1\over2}\big({{4}\over{r}}+\nu{'}-\lambda{'}\big)\big[1+8\pi\xi(6\xi-1)\psi^2\big]
+8\pi\xi(6\xi-1)\psi\psi{'}\Big]
-\mu^2e^{\lambda}\big(1+8\pi\xi\psi^2\big)\psi=0\  ,
\end{equation}
which determines the spatial behavior of the nonminimally coupled
massive scalar fields in the static black-hole spacetime
(\ref{Eq1}).

We note that, in the dimensionless regimes $\xi\geq 1/6$ and
$\xi\leq0$ of the coupling parameter $\xi$, the functional
expression
\begin{equation}\label{Eq25}
{\cal F}(r;\xi)\equiv 1+8\pi\xi(6\xi-1)\psi^2\
\end{equation}
that appears on the l.h.s of (\ref{Eq24}) is a positive definite
function. We shall now prove that ${\cal F}(r)$ is a positive
definite function in the exterior black-hole spacetime also in the
dimensionless physical regime $0<\xi<1/6$. To this end, we first
point out that the characteristic asymptotic behavior (\ref{Eq21})
of the massive scalar fields yields the simple relation [see Eq.
(\ref{Eq25})]
\begin{equation}\label{Eq26}
{\cal F}(r/M\to\infty)\to1\  .
\end{equation}
Let us assume that the function ${\cal F}(r)$ switches signs at some
radial point $r_0\in(r_{\text{H}},\infty)$. Then, the characteristic
differential equation (\ref{Eq24}) yields the simple relation
\cite{Notepno}
\begin{equation}\label{Eq27}
8\pi\xi(6\xi-1)(\psi{'})^2=\mu^2e^{\lambda}(1+8\pi\xi\psi^2)\ \ \ \
\text{at}\ \ \ \ r=r_0\
\end{equation}
for the radial eigenfunction of the nonminimally coupled massive
scalar fields at the assumed root of ${\cal F}(r)$. Noting that the
functional expression on the l.h.s of (\ref{Eq27}) is non-positive
in the physical regime $0<\xi<1/6$ whereas the functional expression
on the r.h.s of (\ref{Eq27}) is positive definite, one deduces that,
in the exterior black-hole spacetime, the function ${\cal F}(r)$
{\it cannot} switch signs. In particular, from the asymptotic
behavior (\ref{Eq26}), one obtains the simple relation
\begin{equation}\label{Eq28}
{\cal F}(r)>0\ \ \ \ \text{for}\ \ \ \ r\in(r_{\text{H}},\infty)\  .
\end{equation}

We further note that one deduces from Eqs. (\ref{Eq2}), (\ref{Eq3}),
(\ref{Eq12}), and (\ref{Eq15}) that the radial eigenfunction
$\psi(x)$ of the nonminimally coupled massive scalar fields is
characterized by the bounded near-horizon functional behavior
\cite{Notefoh}
\begin{equation}\label{Eq29}
|\psi\psi{'}(x\to0)|<\infty\  .
\end{equation}

Taking cognizance of Eqs. (\ref{Eq2}), (\ref{Eq3}), (\ref{Eq28}),
and (\ref{Eq29}), one obtains from (\ref{Eq24}) the near-horizon
($x\ll1$) radial equation
\begin{equation}\label{Eq30}
\psi{''}\cdot\big[1+8\pi\xi(6\xi-1)\psi^2\big]
+\psi{'}\cdot{{1}\over{xr_{\text{H}}}}\big[1+8\pi\xi(6\xi-1)\psi^2\big]
-{{\mu^2}\over{Lx}}\big(1+8\pi\xi\psi^2\big)\psi=0\
\end{equation}
for the nonminimally coupled massive scalar fields. The physically
acceptable solution of the radial scalar equation (\ref{Eq30}) which
respects the relation (\ref{Eq29}) is characterized by the small-$x$
(near-horizon) functional behavior \cite{Notefoh,Noteaq0,Notell2}
\begin{equation}\label{Eq31}
\psi(x\to0)=a\Big[1+{{\mu^2r^2_{\text{H}}}\over{L}}{{1+8\pi\xi
a^2}\over{1+8\pi\xi(6\xi-1)a^2}}\cdot x\Big]+O(x^2)\  ,
\end{equation}
where $a$ is a constant. From Eqs. (\ref{Eq2}), (\ref{Eq28}), and
(\ref{Eq31}) one deduces that, in the $\xi\geq0$ regime, the radial
scalar eigenfunction of the nonminimally coupled massive scalar
fields is characterized by the near-horizon ($x\ll1$) functional
behavior \cite{Noteaq0,Notell2}
\begin{equation}\label{Eq32}
\psi\psi'(x\to0)>0\ .
\end{equation}

Taking cognizance of the analytically derived small-$x$ and
large-$x$ spatial behaviors (\ref{Eq21}) and (\ref{Eq32}), which
characterize the radial eigenfunction $\psi(x)$ of the static
nonminimally coupled massive scalar field configurations, one
deduces that the function $\psi(r)$ has a non-monotonic dependence
on the radial coordinate $r$. In particular, the spatially regular
scalar eigenfunction $\psi(r)$ must have (at least) one extremum
point $r_{\text{peak}}\in(r_{\text{H}},\infty)$ in the exterior
black-hole spacetime which is characterized by the simple functional
relations
\begin{equation}\label{Eq33}
\{\psi\neq0\ \ \ ; \ \ \ \psi{'}=0\ \ \ ; \ \ \
\psi\cdot\psi{''}<0\}\ \ \ \ \text{for}\ \ \ \ r=r_{\text{peak}}\  .
\end{equation}

From Eqs. (\ref{Eq24}) and (\ref{Eq25}) one obtains the functional
relation
\begin{equation}\label{Eq34}
{\cal F}\cdot\psi\psi{''}=\mu^2e^{\lambda}(1+8\pi\xi\psi^2)\psi^2\ \
\ \ \text{at}\ \ \ \ r=r_{\text{peak}}\
\end{equation}
at the extremum point $r=r_{\text{peak}}$ [see Eq. (\ref{Eq33})]
which characterizes the non-monotonic radial eigenfunction of the
nonminimally coupled massive scalar fields. From Eqs. (\ref{Eq28})
and (\ref{Eq33}) one deduces that the radial expression on the l.h.s
of (\ref{Eq34}) is negative definite whereas the radial expression
on the r.h.s of (\ref{Eq34}) is positive definite in the
dimensionless physical regime $\xi\geq0$. One therefore realizes
that the functional relation (\ref{Eq34}), which characterizes the
spatial behavior of the non-monotonic radial scalar eigenfunction at
the extremum point (\ref{Eq33}), {\it cannot} be respected. We
therefore conclude that, in the dimensionless regime $\xi\geq0$ of
the nonminimal field-curvature coupling parameter $\xi$,
asymptotically flat static matter configurations made of
nonminimally coupled massive scalar fields cannot be supported by
spherically symmetric black holes with regular event horizons.

\section{Summary}

The physically interesting no-hair theorems of Bekenstein and Mayo
\cite{BekMay,Bek20} can be used to rule out the existence of
spherically symmetric neutral black holes with regular event
horizons that support static massive scalar fields nonminimally
coupled to gravity in the dimensionless regimes $\xi\leq0$ and
$\xi\geq 1/2$ \cite{Noteev}. The no-hair theorem presented in
\cite{Hodrs1} can be used to extend the validity of the no-hair
property to the dimensionless regime $\xi>1/4$. Intriguingly,
however, no mathematically rigorous theorem has so far been
presented in the physics literature which rules out the existence of
asymptotically flat static hairy black-hole-massive-scalar-field
configurations in the dimensionless complementary regime
$0<\xi\leq1/4$.

Studying analytically the non-linearly coupled Einstein-scalar field
equations, we have explicitly proved in the present paper that
spherically symmetric asymptotically flat neutral black holes with
regular event horizons cannot support static matter configurations
made of nonminimally coupled massive scalar fields with {\it
generic} values of the dimensionless physical parameter $\xi$.


\bigskip
\noindent
{\bf ACKNOWLEDGMENTS}
\bigskip

This research is supported by the Carmel Science Foundation. I would
like to thank Yael Oren, Arbel M. Ongo, Ayelet B. Lata, and Alona B.
Tea for helpful discussions.



\begin{thebibliography}{99}

\bibitem{Bek1} J. D. Bekenstein, Phys. Rev. Letters {\bf 28}, 452 (1972).

\bibitem{Teit} C. Teitelboim, Lett. Nuov. Cim. {\bf 3}, 326 (1972).

\bibitem{Chas} J. E. Chase, Commun. Math. Phys. {\bf 19}, 276 (1970).

\bibitem{Bekto} J. D. Bekenstein, Physics Today {\bf 33}, 24 (1980).

\bibitem{Heu} M. Heusler, J. Math. Phys. {\bf 33}, 3497 (1992);
M. Heusler, Class. Quant. Grav. {\bf 12}, 779 (1995).

\bibitem{Bek2} J. D. Bekenstein, Phys. Rev. D {\bf 51}, R6608 (1995).

\bibitem{BekMay} A. E. Mayo and J. D. Bekenstein, Phys. Rev. D {\bf 54}, 5059 (1996).

\bibitem{Bek20} J. D. Bekenstein, arXiv:gr-qc/9605059 .

\bibitem{Noteregn} The term `regular black hole' is used here to
describe a black-hole spacetime with a non-singular boundary (event
horizon).

\bibitem{Noteev} It is worth mentioning that, as explicitly proved in
\cite{Bek20,BekMay}, the intriguing no-hair property of
asymptotically flat spherically symmetric black-hole spacetimes with
regular event horizons can be extended to the case of scalar fields
with positive semidefinite self-interaction potentials
\cite{Bek20,BekMay}.

\bibitem{Hodrc} S. Hod, Phys. Rev. D {\bf 86}, 104026 (2012) [arXiv:1211.3202];
S. Hod, The Euro. Phys. Journal C {\bf 73}, 2378 (2013)
[arXiv:1311.5298]; S. Hod, Phys. Rev. D {\bf 90}, 024051 (2014)
[arXiv:1406.1179]; S. Hod, Phys. Lett. B {\bf 739}, 196 (2014)
[arXiv:1411.2609]; S. Hod, Class. and Quant. Grav. {\bf 32}, 134002
(2015) [arXiv:1607.00003]; S. Hod, Phys. Lett. B {\bf 751}, 177
(2015); S. Hod, Class. and Quant. Grav. {\bf 33}, 114001 (2016); S.
Hod, Phys. Lett. B {\bf 758}, 181 (2016) [arXiv:1606.02306]; S. Hod
and O. Hod, Phys. Rev. D {\bf 81}, 061502 Rapid communication (2010)
[arXiv:0910.0734]; S. Hod, Phys. Lett. B {\bf 708}, 320 (2012)
[arXiv:1205.1872]; S. Hod, Jour. of High Energy Phys. {\bf 01}, 030
(2017) [arXiv:1612.00014].

\bibitem{Herkr} C. A. R. Herdeiro and E. Radu, Phys. Rev. Lett. {\bf 112}, 221101
(2014); C. L. Benone, L. C. B. Crispino, C. Herdeiro, and E. Radu,
Phys. Rev. D {\bf 90}, 104024 (2014); C. A. R. Herdeiro and E. Radu,
Phys. Rev. D {\bf 89}, 124018 (2014); C. A. R. Herdeiro and E. Radu,
Int. J. Mod. Phys. D {\bf 23}, 1442014 (2014); Y. Brihaye, C.
Herdeiro, and E. Radu, Phys. Lett. B {\bf 739}, 1 (2014); J. C.
Degollado and C. A. R. Herdeiro, Phys. Rev. D {\bf 90}, 065019
(2014); C. Herdeiro, E. Radu, and H. R\'unarsson, Phys. Lett. B {\bf
739}, 302 (2014); C. Herdeiro and E. Radu, Class. Quantum Grav. {\bf
32} 144001 (2015); C. A. R. Herdeiro and E. Radu, Int. J. Mod. Phys.
D {\bf 24}, 1542014 (2015); C. A. R. Herdeiro and E. Radu, Int. J.
Mod. Phys. D {\bf 24}, 1544022 (2015); P. V. P. Cunha, C. A. R.
Herdeiro, E. Radu, and H. F. R\'unarsson, Phys. Rev. Lett. {\bf
115}, 211102 (2015); B. Kleihaus, J. Kunz, and S. Yazadjiev, Phys.
Lett. B {\bf 744}, 406 (2015); C. A. R. Herdeiro, E. Radu, and H. F.
R\'unarsson, Phys. Rev. D {\bf 92}, 084059 (2015); C. Herdeiro, J.
Kunz, E. Radu, and B. Subagyo, Phys. Lett. B {\bf 748}, 30 (2015);
C. A. R. Herdeiro, E. Radu, and H. F. R\'unarsson, Class. Quant.
Grav. {\bf 33}, 154001 (2016); C. A. R. Herdeiro, E. Radu, and H. F.
R\'unarsson, Int. J. Mod. Phys. D {\bf 25}, 1641014 (2016); Y.
Brihaye, C. Herdeiro, and E. Radu, Phys. Lett. B {\bf 760}, 279
(2016); Y. Ni, M. Zhou, A. C. Avendano, C. Bambi, C. A. R. Herdeiro,
and E. Radu, JCAP {\bf 1607}, 049 (2016); M. Wang, arXiv:1606.00811
.

\bibitem{NR1} See also Y. Shlapentokh-Rothman, Commun. Math. Phys. {\bf 329}, 859 (2014);
G. Moschidis, arXiv:1608.02041 .

\bibitem{Notewx} The physical parameter $\xi$ quantifies
the strength of the non-trivial coupling between the massive scalar
field and the spacetime curvature [see Eq. (\ref{Eq5}) below].

\bibitem{Noters} See \cite{Hodrs1,Hodrs2} for novel no-scalar-hair
theorems for spherically symmetric asymptotically flat reflecting
(rather than absorbing) compact objects.

\bibitem{Hodrs1} S. Hod, Phys. Rev. D {\bf 94}, 104073 (2016) [arXiv:1612.04823];
S. Hod, Phys. Rev. D {\bf 96}, 024019 (2017) [arXiv:1709.01933]; S.
Hod, Phys. Lett. B {\bf 773}, 208 (2017).

\bibitem{Hodrs2} S. Hod, Phys. Lett. B {\bf 771}, 521 (2017).

\bibitem{NRs} T. P. Sotiriou, Class. Quant. Grav. {\bf 32}, 214002
(2015) [arXiv:1505.00248].

\bibitem{NR2} See also T. P. Sotiriou and V. Faraoni, Phys. Rev. Lett. {\bf 108}, 081103
(2012); L. Hui and A. Nicolis, Phys. Rev. Lett. {\bf 110}, 241104
(2013); T. P. Sotiriou and S. Y. Zhou, Phys. Rev. Lett. {\bf 112},
251102 (2014).

\bibitem{NR3} See also M. Salgado, Class. Quant. Grav. {\bf 23}, 4719 (2006);
M. Salgado, D. Martinez-del Rio, M. Alcubierre and D. Nunez, Phys.
Rev. D {\bf 77}, 104010 (2008).

\bibitem{Asp} R. L. Bowers and E. P. T. Liang, The Astrophys. Jour., {\bf 188}, 657
(1974).

\bibitem{Notepno} Note that $\psi\neq0$ at the assumed root of
${\cal F}$ [see Eq. (\ref{Eq25})].

\bibitem{Notefoh} It is worth emphasizing that one learns
from Eqs. (\ref{Eq2}), (\ref{Eq3}), and (\ref{Eq24}) that if ${\cal
F}(x)$ vanishes in the near-horizon $x\to0$ limit, then
$\psi\psi'(x\to0)$ blows up. This divergent near-horizon behavior
contradicts the bounded functional relation (\ref{Eq29}) [see
(\ref{Eq22})]. One therefore concludes that ${\cal F}(x)$ cannot
vanish in the near-horizon $x\to0$ limit [see also (\ref{Eq28})].

\bibitem{Noteaq0} Note that $a\equiv\psi(x\to0)\neq0$ [see Eq.
(\ref{Eq22})]. As emphasized above, if $\psi(x\to0)=0$ then the
functional relation (\ref{Eq21}), which characterizes the asymptotic
spatial behavior of the radial eigenfunction of the massive scalar
fields, implies that $\psi(r)$ must have an extremum point in the
exterior black-hole spacetime in accord with our previous assertion.

\bibitem{Notell2} Note that, as proved in \cite{BekMay}, $L>0$.


\end{thebibliography}
\end{document}